\documentclass{article}

\usepackage{arxiv}

\usepackage[utf8]{inputenc} 
\usepackage[T1]{fontenc}    
\usepackage{hyperref}       
\usepackage{url}            
\usepackage{booktabs}       
\usepackage{amsfonts}       
\usepackage{nicefrac}       
\usepackage{microtype}      
\usepackage{lipsum}		
\usepackage{graphicx}
\usepackage{natbib}
\usepackage{doi}
\usepackage{comment}
\usepackage{amsmath}
\usepackage{chemformula}
\usepackage{caption}
\usepackage{subcaption}
\usepackage{multirow}

\title{Reducing the Environmental Impact of Midblock Crossing}

\author{{ Abrar Alali}\\
	Department of Computer Science\\
	Old Dominion University\\
	Norfolk, VA 23529 \\
	\texttt{aalal003@odu.edu} \\
	\And
	{Stephan Olariu} \\
	Department of Computer Science\\
	Old Dominion University\\
	Norfolk, VA 23529 \\
	\texttt{solariu@odu.edu} \\
}

\renewcommand{\shorttitle}{\textit{arXiv} Template}

\hypersetup{
pdftitle={Reducing the Environmental Impact of Midblock Crossing},
pdfsubject={q-bio.NC, q-bio.QM},
pdfauthor={Abrar Alali, Stephan Olariu},
pdfkeywords={Smart mobility , Fuel consumption , \ch{CO2} Emissions , Vehicular communications , Midblock crossing , Pedestrian safety},
}

\begin{document}
\maketitle

\begin{abstract}
	Accommodating pedestrians crossing midblock has been shown to have harmful environmental consequences because of increased fuel consumption and \ch{CO2} emissions. 
Somewhat surprisingly, no studies were devoted to mitigating the environmental impact of midblock crossing.
Our main contribution is to propose schemes that mitigate the increased fuel consumption and \ch{CO2} emissions due to pedestrian midblock crossing by leveraging 
information about the location and expected duration of the crossing. This information is shared, in a timely manner, with approaching cars. We evaluated the impact of car decisions 
on fuel consumption and emissions by exploring potential trajectories that cars may take as a result of messages received. Our extensive simulations showed that timely dissemination 
of pedestrian crossing information to approaching vehicles can reduce fuel consumption and emissions by up to 16.7\%
\end{abstract}

\keywords{Smart mobility \and Fuel consumption \and \ch{CO2} Emissions \and Vehicular communications \and Midblock crossing \and Pedestrian safety}

\section{Introduction}
\label{sec:intro}

Recent statistics revealed that midblock crossing, including jaywalking,  is a ubiquitous and pervasive societal phenomenon that is here to stay~\cite{HUNSANON20171672,Tezcan2019-su}. 
Further studies have confirmed that accommodating pedestrians who cross midblock increases fuel consumption, \ch{CO2} emissions, and the average trip time 
\cite{Li2014-jq,app112411794,Bak2012-ba}. 
In order to avoid crashing into pedestrians crossing midblock, cars must reduce their speed and then accelerate to resume their cruising speed. Unfortunately, 
these avoidance maneuvers increase significantly fuel consumption and emissions~\cite{El-Shawarby2005-lm}. 

The common approach taken by researchers to contain the increase in fuel consumption due to promoting pedestrian safety at controlled intersections involves using 
vehicle-to-infrastructure (V2I) communications (e.g., intersection management system) and/or vehicle-to-vehicle (V2V) communications protocols
~\cite{Alsabaan2013-zt, Wan2016-po, Lu2019-cp, Wang2021-cl}. If pedestrians are detected at intersections managed by traffic systems, the approaching cars react according 
to the scheduling scheme sent by the traffic management system.

Pedestrians who cross midblock can be detected using on-board pedestrian detection systems~\cite{Palffy2023-oz} or by collaborative perception 
of surrounding cars and/or infrastructure~\cite{ALALI2023100601, Zhang2022-ws, noh2022novel}. 

\subsection{Our contributions}

Virtually all studies on reducing the environmental impacts of promoting road safety
have focused on communication between intersection management systems and vehicles to optimize fuel consumption. However, as previously mentioned, cars can
now receive alerts about midblock crossings. Therefore, developing methods to efficiently reduce the speed based on the received alert messages is required.

The main contribution of this paper is to fill this research gap by providing schemes for adjusting the speed of the car after receiving midblock crossing alerts considering the 
environmental impacts. While environmental impacts of midblock crossing have been explored, there is a glaring lack of effort to reduce these impacts. 
We propose two schemes for maintaining a safe speed to avoid collisions with pedestrians crossing at several different locations, without the need for the cars to stop. 
The first scheme is to immediately reduce the speed to a safe speed that minimizes fuel consumption and reduces emissions. The second scheme is to defer the deceleration if the 
car is already responding to a previous alert until it reaches the crossing area.

Our extensive simulation results show that timely dissemination of pedestrian crossing information to approaching cars can reduce fuel
consumption and emissions by up to 16.7\% 

The remainder of this paper is organized as follows. In Sections \ref{sec:rel-work} and \ref{sec:fuel-and-emissions} we review, succinctly, related work and 
the fuel consumption model used. We present our alert model in Section~\ref{sec:sys-model}. In Section~\ref{sec:method} we illustrate our schemes for reducing speed upon 
receiving alert messages. We show the results of implementing 
our proposed schemes in Section~\ref{sec:results}. Finally, we offer concluding remarks in Section~\ref{sec:conc}.

\section{Literature review}\label{sec:rel-work}
A number of researchers have utilized information exchanged through vehicular networks to reduce the environmental impact of transportation systems. 
For example, ~\cite{Alsabaan2013-zt} introduced a comprehensive optimization model involving both V2V and traffic-light-signal-to-vehicle (TLS2V) communications. 
Through V2V and V2I communications, cars approaching a traffic light signal receive information to adjust their speed to a recommended value, aiming to minimize fuel 
consumption and emissions. Similarly,~\cite{Wan2016-po} proposed a Speed Advisory System (SAS) for connected vehicles to enhance fuel efficiency and comfort by managing speed 
in advance based on upcoming traffic signal information.
Further,  \cite{Lu2019-cp} introduced an advanced speed control at successive signalized intersections to mitigate fuel consumption and emissions, leveraging V2I and V2V 
technologies. Car speed was optimized by utilizing real-time traffic signal phasing, timing information, and vehicle queue data. The method notably reduced fuel consumption and 
\ch{CO2} emissions by over 18\%. In addition, pedestrians crossing intersections were accommodated in \cite{Wang2021-cl} who introduced Roadrunner+, a cooperative autonomous 
intersection management system designed for connected autonomous vehicles. This work addressed the challenges posed by pedestrian crossings at intersections and scheduled 
traffic efficiently reducing fuel consumption by up to 7.64\%.

\section{Fuel consumption and emission estimation model}\label{sec:fuel-and-emissions}
The physics-based energy demand model was used by \citet{Jones1980-qt} to estimate the energy required to move a car from point to point through a driving cycle. The model takes as 
inputs the speed profile and car specifications that influence the tractive force. We employed the model to estimate fuel consumption based on the instantaneous energy 
demand using the following equation~\citet{thomas2014drive}:
\begin{equation}
E_{Fuel} = \frac{E_{inst}}{\eta},
\end{equation}
\noindent
where $E_{Fuel}$ is the estimation of the total expended fuel energy in $Joules$ and $\eta$ refers to the efficiency of the engine in converting fuel into a tractive power. The total instantaneous energy demand $E_{inst}$ on the car in the interval $[0,T]$ is calculated as follows:
\begin{equation}
E_{inst} = \sum_{i=0}^{T}~m~a_i~v_i + f_0~v_i + f_2~v_i^3,
\end{equation}
\noindent
where $i$ is the time unit (i.e., one second), $m$ is the mass of the car in $kg$, $a_i$ is the current acceleration of the 
car in $m/sec^2$ and $v_i$ is the current speed in $m/sec$. $f_0$,$f_2$ are the rolling resistance and aerodynamic drag of the car, respectively. These terms represent the 
forces that a car must overcome while moving \citep{guzzella2007vehicle}. 
As stated by \cite{fueleconomy_atv} (EPA), only 12\% to 30\% of the fuel goes to the wheel of the car. We assume that the power demand during deceleration is zero because the 
engine does not provide power to the wheels while braking. Similarly, the power demand during idling is zero as the car shuts down the engine if it stops for more 
than a few seconds  \citep{huff2023auto}. 

To estimate the \ch{CO2} emissions, we use the EPA formula \cite{epa_EII}:
\begin{equation}
\ch{CO2} = E_{Fuel} \times  \text{Carbon Content} \times \text{Oxidation Fraction} \times \left(\frac{44}{12}\right),
\end{equation}
\noindent
\ch{CO2} emissions are highly correlated with the fuel consumed. Based on this, the \ch{CO2} emissions in grams per $KJoules$ can be estimated based on the expended fuel 
energy $E_{Fuel}$. The Carbon Content equals to 0.0196 g/KJ and the Oxidation Fraction equals to 0.99 and the value $\frac{44}{12}$ is the  molecular mass of \ch{CO2} 
divided by the atomic mass of carbon.

\section{The alert system model}\label{sec:sys-model}

\begin{figure}[ht!]
\centering
\includegraphics[width=0.6\textwidth]{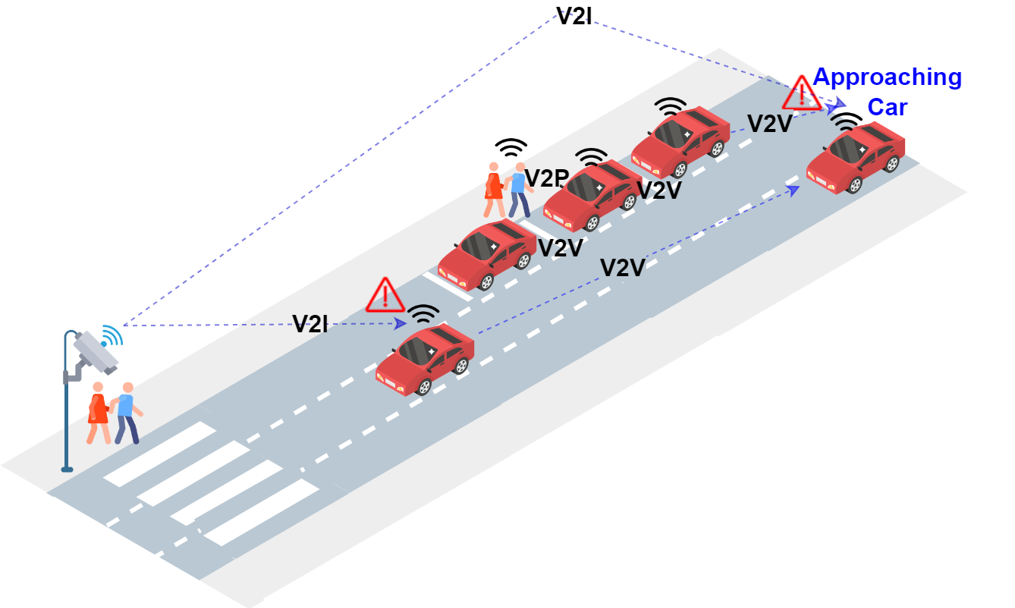}
\caption{\em A generic alert system model.}
\label{fig:Systemmodel}

\end{figure}

We assume a system that can detect reliably pedestrians crossing midblock and that sends alert messages to approaching cars, as illustrated  in Fig. ~\ref{fig:Systemmodel}. For example, 
approaching cars may receive, through V2I communications, alert messages from a street monitoring system that uses cameras~\cite{noh2022novel}. It may also receive, 
through V2V communications, alert messages from cars parked along the curb that detect crossing pedestrians~\cite{ALALI2023100601}. Similarly, the approaching cars may receive 
alert messages from other approaching cars through V2V~\cite{Ngo2023-lm}. Finally, cars may receive alert messages from pedestrian hand-held or wearable devices 
through Vehicle-to-Pedestrian (V2P) communications~\cite{Tahmasbi-Sarvestani2017-ym}. 

We assume that alert messages include the location of the crossing and the speed of the pedestrians. Using a digital map, the car determines the width of the street. 
With this information, the car can determine the remaining crossing time and the speed it should maintain to avoid colliding with the pedestrians, as we illustrate 
in Section~\ref{sec:method}.



\section{Speed reduction schemes}\label{sec:method}

\begin{figure}
     \centering
     \begin{subfigure}[b]{0.4\textwidth}
         \centering
         \includegraphics[width=\textwidth]{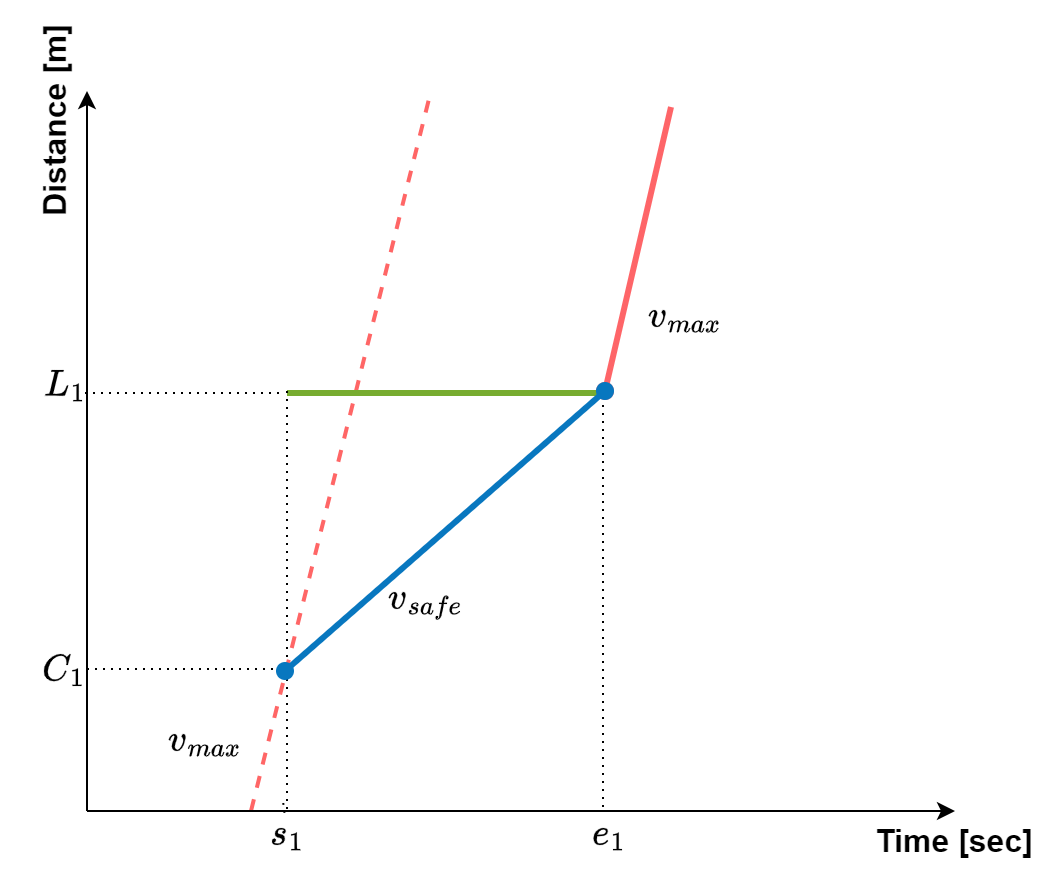}
         \caption{One Alert}
         \label{fig:oneped}
     \end{subfigure}
     \begin{subfigure}[b]{0.4\textwidth}
        \centering
        \includegraphics[width=\textwidth]{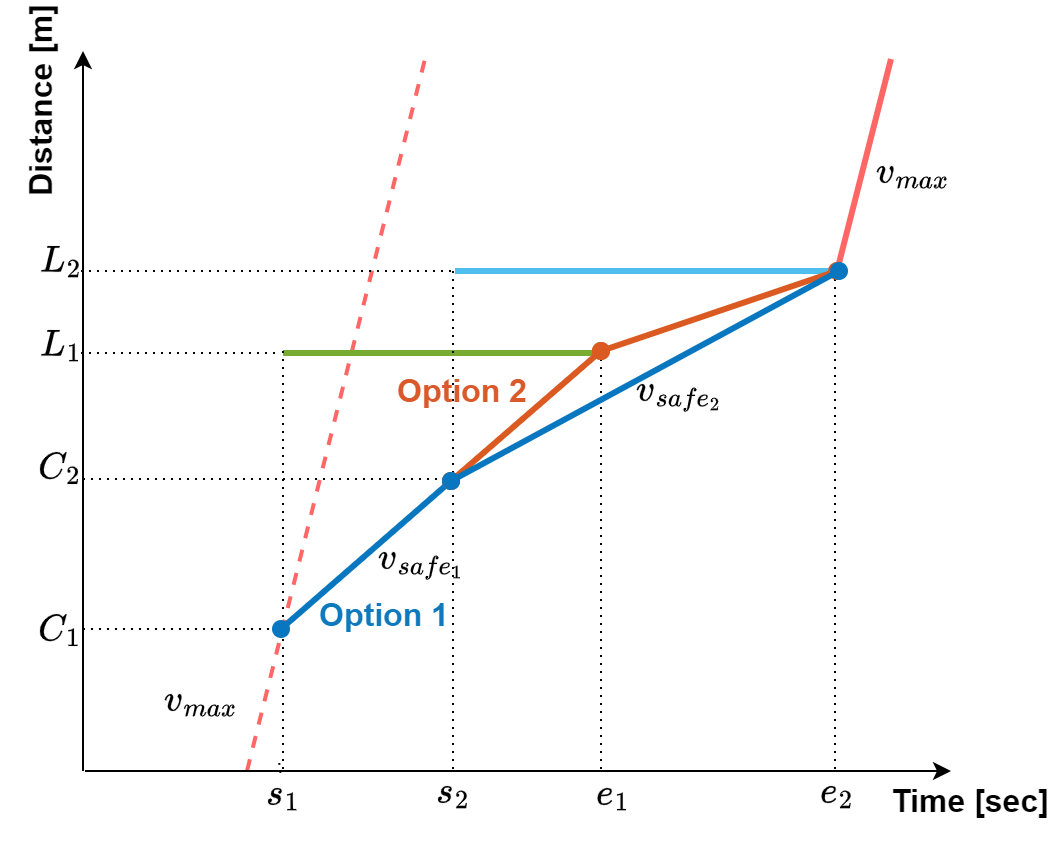}
        \caption{Two Alerts}
        \label{fig:multi-ped}
     \end{subfigure}
    \caption{Trajectory of a car receiving alert message about midblock crossing}
    \label{fig:drivingcycles}
  
\end{figure}

Assume that at time $s_1$ an approaching car at location $C_1$ receives the first alert message containing location $L_1$ for the crossing cohort and the remaining crossing time, 
which indicates that the crossing will be over at time $e_1$. Using the time-space diagram shown in Fig. ~\ref{fig:oneped}, the car can calculate the maximum safe speed as:
\begin{equation}
	v_{safe_i} = \min \left \{ v_{safe_{i-1}} , \frac{L_i-C_i}{e_i-s_i} \right \},
\end{equation}
where $i$ is the alert message number. $v_{safe_i}$ is the safe speed determined using the information included in the received message number $i$. If the received message is the first message, then $v_{safe_{i}}$ is equal to the street speed limit $v_{max}$.


Now, let us assume that after a while and before the car reaches the crossing location $L_1$ of the first crossing, it receives another message about a cohort crosses at different location $L_2$, as we show in Fig.~\ref{fig:multi-ped}.
When the car receives the alert $2$, it checks whether there is a previous alert $1$ affects its speed. If no alert was received, the car reduces its speed immediately according to the calculated crossing time. On the other hand, if there is a previous alert, the car will check if the location of the current crossing pedestrian $L_{2}$ is closer than that of the previous pedestrian location $L_1$. If it is closer, then the car would choose between the two options, which represent our proposed schemes, Option 1 or Option 2, as explained below: 
\noindent

\begin{itemize}
     \item \textbf{Option 1: Immediate deceleration}
        This option is illustrated in Fig. ~\ref{fig:multi-ped}. In this option, the car at location $C_2$ responds immediately to the second message by reducing its speed. We note here that this option would also prevent collisions with pedestrians at location $L_1$ without the need to stop at $L_1$. Indeed, when the car reaches the crossing area, it can resume a normal speed equal to the speed limit $v_{max}$.
    
    \item \textbf{Option 2: Deferred deceleration:}
        This option represents the second scheme we propose in this work. This option is depicted in Fig. ~\ref{fig:multi-ped}. In this option, after the car reduces its speed for crossing at location $L_1$, when it is at location $C_2$ it receives the second message about cohort crossing at location $L_2$. When the car compares the new safe speed $v_{safe_2}$ with its current speed $v_{safe_1}$, it finds that $v_{safe_1}$ is lower. Therefore, it may choose not to respond to this message and maintain its current speed until it reaches the crossing at $L_1$. When it passes the area, it will reduce its speed according to the remaining crossing time for the cohort at location $L_2$.
    \end{itemize}

These two proposed schemes can be compared with other trajectories that can be taken when the car does not notice the pedestrian in advance. We call this trajectory (Sudden Stop) where there are midblock crossings, but the car does not receive any information beforehand, and alternatively it notices the pedestrian near the crossing location and stops suddenly. Note that the (Sudden Stop) trajectory is similar to a trajectory that would be taken by an aggressive driver who does not adhere to the alert message.

All of these trajectories can be compared to a baseline trajectory when there are no midblock crossing pedestrians. We call this trajectory (No Peds.) which stands for No Pedestrians.

\begin{figure}
     \centering
     \begin{subfigure}[b]{0.25\textwidth}
         \centering
         \includegraphics[width=\textwidth]{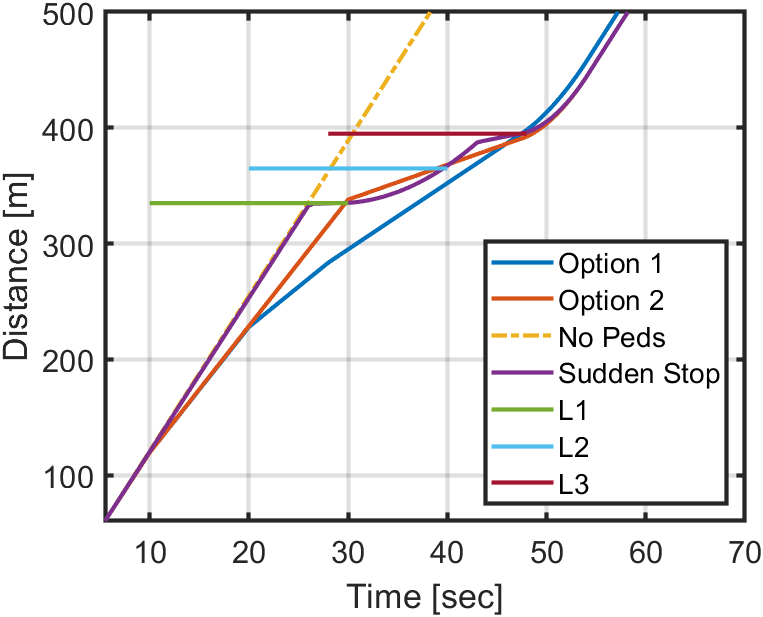}
         \caption{Scenario 1}
         \label{fig:sce1}
     \end{subfigure}
     \hspace{0.3cm}
     \begin{subfigure}[b]{0.25\textwidth}
        \centering
        \includegraphics[width=\textwidth]{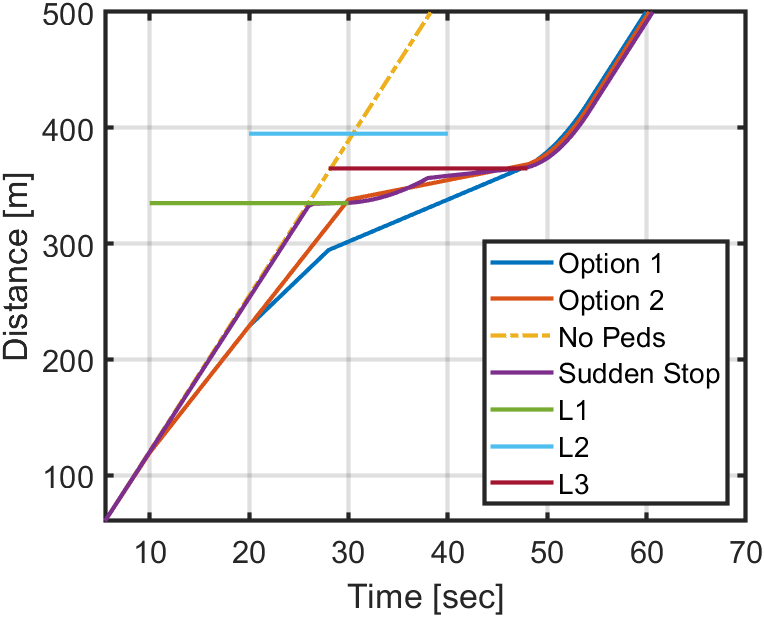}
        \caption{Scenario 2}
        \label{fig:sce2}
     \end{subfigure}
     \hspace{0.3cm}
     \begin{subfigure}[b]{0.25\textwidth}
         \centering
         \includegraphics[width=\textwidth]{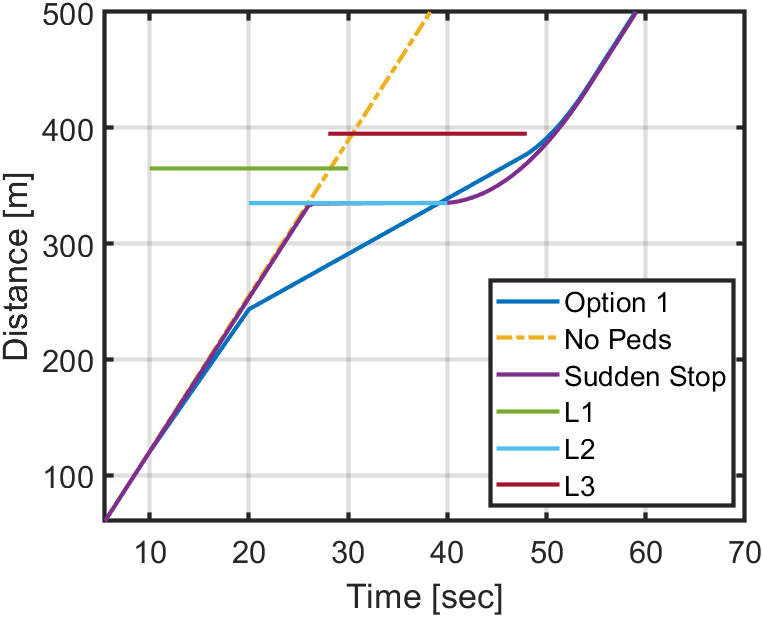}
         \caption{Scenario 3}
         \label{fig:sce3}
     \end{subfigure}

     \begin{subfigure}[b]{0.25\textwidth}
        \centering
        \includegraphics[width=\textwidth]{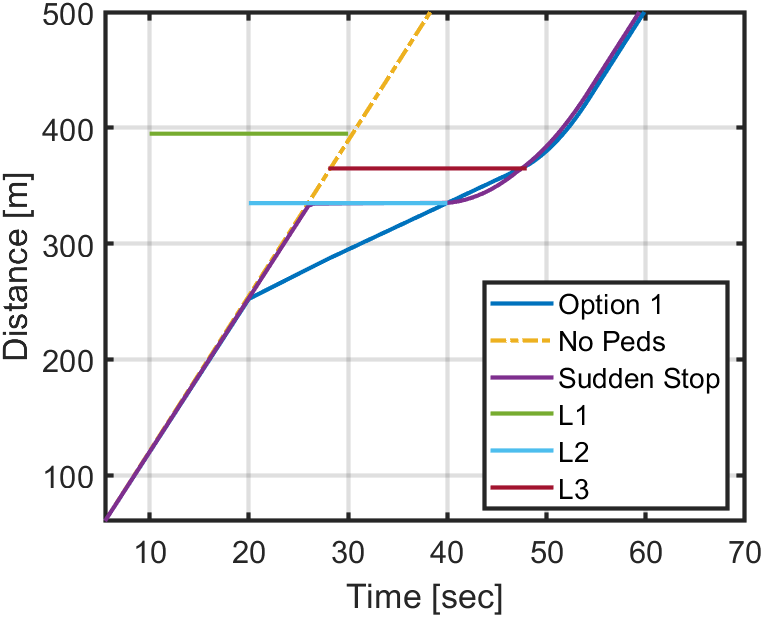}
        \caption{Scenario 4}
        \label{fig:sce4}
     \end{subfigure}
     \hspace{0.3cm}
     \begin{subfigure}[b]{0.25\textwidth}
         \centering
         \includegraphics[width=\textwidth]{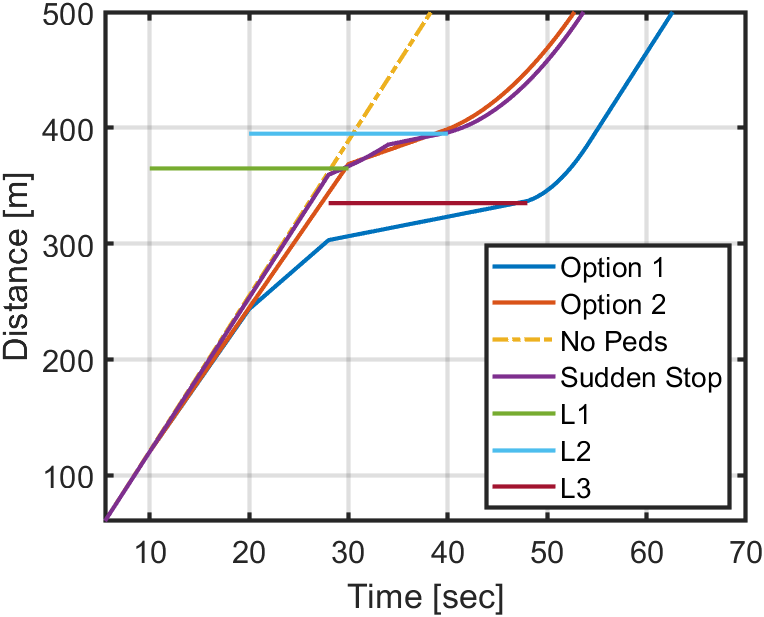}
         \caption{Scenario 5}
         \label{fig:sce5}
     \end{subfigure}
     \hspace{0.3cm}
     \begin{subfigure}[b]{0.25\textwidth}
        \centering
        \includegraphics[width=\textwidth]{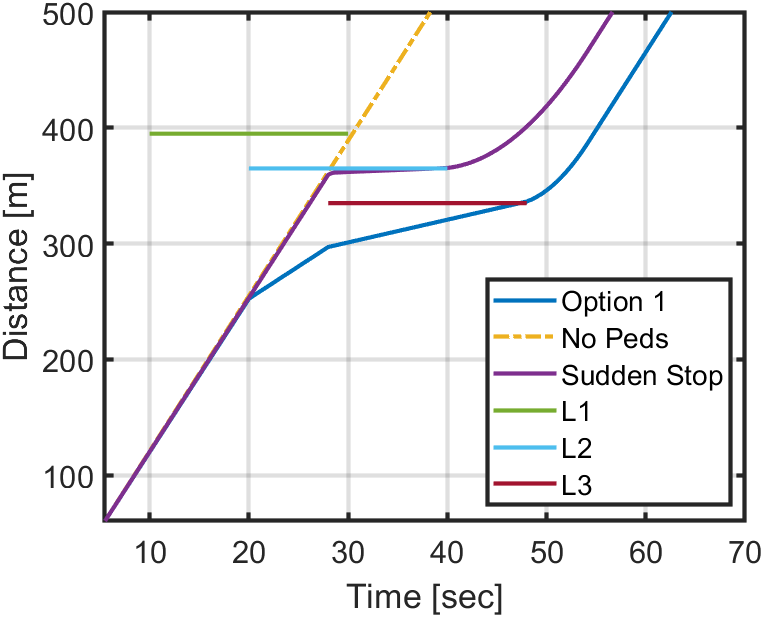}
        \caption{Scenario 6}
        \label{fig:sce6}
     \end{subfigure}

    \caption{Evaluated scenarios of vehicle trajectory when receiving three alert messages}
    \label{fig:scenarios}

\end{figure}

It is important to determine the possible scenarios that may occur when there are several simultaneous crossings at several locations, and an approaching car receives alert messages. This also enables an accurate evaluation of the effectiveness of each reaction in reducing the environmental impacts. Therefore, we designed six possible scenarios for three pedestrians crossing at the same speed (i.e., they had the same crossing time) at three different locations L1, L2, and L3. In each scenario, the order of starting crossing changes at each location. Fig.~\ref{fig:scenarios} shows six scenarios with their expected reactions.

In scenarios 3, 4 and 6, (Fig.~\ref{fig:sce3}, \ref{fig:sce4}, and \ref{fig:sce6}), the only option for the car to avoid collisions with pedestrians without staying idle is immediate deceleration (that is, Option 1). This is the only available option, given that the second alert location is closer to the car than the first alert. In Scenario 3, the car receives the message at time 10 about a pedestrian/cohort crosses at location $L_1$, the car calculates the safe speed $v_{safe_{1}}$ and finds that it is equal to the speed limit, so it continues cruising at the same speed. Note that even if the car does not reduce its speed, it will avoid the collision with the pedestrian without the need to stop at the crossing location. At time 20, the car receives a second alert for a crossing at location $L_2$. Now, because the location is closer than $L_1$ and $v_{safe_{2}}$ is less than the current speed $v_{safe_{1}}$, the car immediately reduces its speed to $v_{safe_{2}}$. When the third crossing starts at location $L_3$ which is farther away from $L_2$, the car calculates $v_{safe_{3}}$ which, this time, is larger than the current speed, so it keeps the minimum which is the current speed $v_{safe_{2}}$. When it reaches $L_2$, its speed can be increased to $v_{safe_{2}}$ if it is less than the speed limit.
In Scenario 4, crossing at location $L_1$ starts first, and the car receives the first alert, but as in Scenario 3, this does not affect the current speed because the safe speed now is similar to the speed limit. At time 20, crossing at a closer location $L_2$ starts, which makes the car reduces its speed to avoid the collision. At time 28, a crossing starts at location $L_3$, but it has no affects this time because $v_{safe_{2}}$ = $v_{safe_{3}}$. 
In scenario 6, the car does not reduce its speed for crossing at $L_1$, but it does so when crossing at location $L_2$ starts. Again, when $L_3$ has a crossing, the car reduces its speed to avoid the collision.

On the other hand, a second option is available to the car to defer the deceleration in Scenarios 1,2 and 5 (Fig. ~\ref{fig:sce1}, \ref{fig:sce2}, and \ref{fig:sce5}). This is actually because the second alert is for crossing at location farther away than the location of the first alert.  
In scenario 1, the car receives the message at time 10 about a pedestrian/cohort crosses at location $L_1$, the car calculates the safe speed and finds that it is less than the speed limit $v_{max}$; thus, it reduces its speed to $v_{safe_{1}}$. Then, when a crossing starts at location $L_2$ which is farther away from $L_1$, it receives the second alert. Now, as the car calculates its $v_{safe_{2}}$ and finds that it is less than the current speed $v_{safe_{1}}$, so it has now two options: either to decelerate immediately to reach $v_{safe_{2}}$ or to continue on $v_{safe{1}}$ until it reaches $L_1$ crossing then decelerates for the second alert. When the third crossing starts at location $L_3$, the car calculates $v_{safe_{3}}$ which, this time, is less than the current speed (for both options), but ,again, since $L_3$ is farther than $L_2$, it has now the two options. 
Similarly, in scenario 2, crossing at location $L_1$ starts first, and the car receives the first alert, then crossing starts at $L_2$, so the car has the two options. However, when crossing at $L_3$ starts, the car must reduce its speed immediately without having a second option because $L_3$ is closer than $L_2$. 
In scenario 5, the car does not reduce its speed for $L_1$ crossing as it can avoid the collision even it maintains the speed limit. Again, when crossing at further location $L_2$ starts, it has the two options. However, when the third crossing at location $L_3$ starts, it affects Option 1 and forces the car to reduce its speed immediately.



In the following, we explain the possible rational reactions of an approaching car inside the Safety Zone when it receives a alert message to a crossing cohort at one location, and after a while, it receives another message of a pedestrian crossing at the same or another location. We examine two cases that may occur in the street while pedestrians cross the midblock.

In this case, the car receives a second alert message about cohort crossing at a location that is different from the first one. 
         
\section{Simulation and results}\label{sec:results}

\subsection{Simulation model}
To generate cars traffic, we utilize Simulation of Urban MObility (SUMO)~\cite{SUMO2018}, a microscopic simulation modeling vehicles and pedestrian mobility. In the simulation, we created a one way street with two lanes for on-street parking in the right-hand of the street and a left-hand lane for moving cars. 

For generating pedestrian traffic, and because we have specific scenarios we aim to study, we determine fixed locations of pedestrian crossing with fixed speed. We chose the minimum speed of crossing midblock reported by~\cite{Forde2021-ge} to assure safety of pedestrians. We assume that each time a car approaches those determined locations, there is a pedestrian crosses the street at the same time, and it receives a message once it enters the Safety Zone. To achieve this, we disable the randomness in pedestrian generation in SUMO to prove the effectiveness of our schemes. The parameters of the simulation are as follows: street length = 500 meters [m], street width = 13 [m], street speed limit = 30 [mph] ~ 13.4 [m/s], pedestrian speed = 0.67 [m/s], crossing time = 20 [s].  

We implemented our proposed schemes on two car models from year 2023 data provided by~\cite{epa_fuel_economy_data}. 
The first car is the sedan TOYOTA CAMRY LE/SE has mass = 1644 kilogram, $f_0$ = 113.82 newton and $f_2$ = 0.36 newton$\times \frac{second^2}{meter^2}$. The second car is the SUV TOYOTA HIGHLANDER that has mass = 2040.8 kg,  $f_0$ = 139.7 newton and $f_2$ = 0.56 newton$\times \frac{second^2}{meter^2}$. According to the EPA data, CAMRY LE/SE  consumes less fuel and emit less \ch{CO2} than HIGHLANDER in city driving cycles.

\subsection{Simulation results}
In this subsection, we provide an evaluation of the proposed schemes in terms of reducing fuel consumption and \ch{CO2} emissions using the two cars. Additionally, we prove that 
receiving informative messages about midblock pedestrians in timely manner via V2V communications reduces the environmental impacts associated with midblock crossing. 
We compare the fuel consumed during the trips in Option 1 and Option 2 in all the assumed scenarios with the additional trajectories (No Peds.) and and (Sudden Stop). Fig.~\ref{fig:scenarios} 
shows the six scenarios 
along with the additional generated trajectories for comparison. We evaluate the fuel consumed and the emitted \ch{CO2} for the four trajectories in each scenario. 

\begin{figure}
     \centering
     \begin{subfigure}[b]{1\textwidth}
         \centering
         \includegraphics[width=\textwidth]{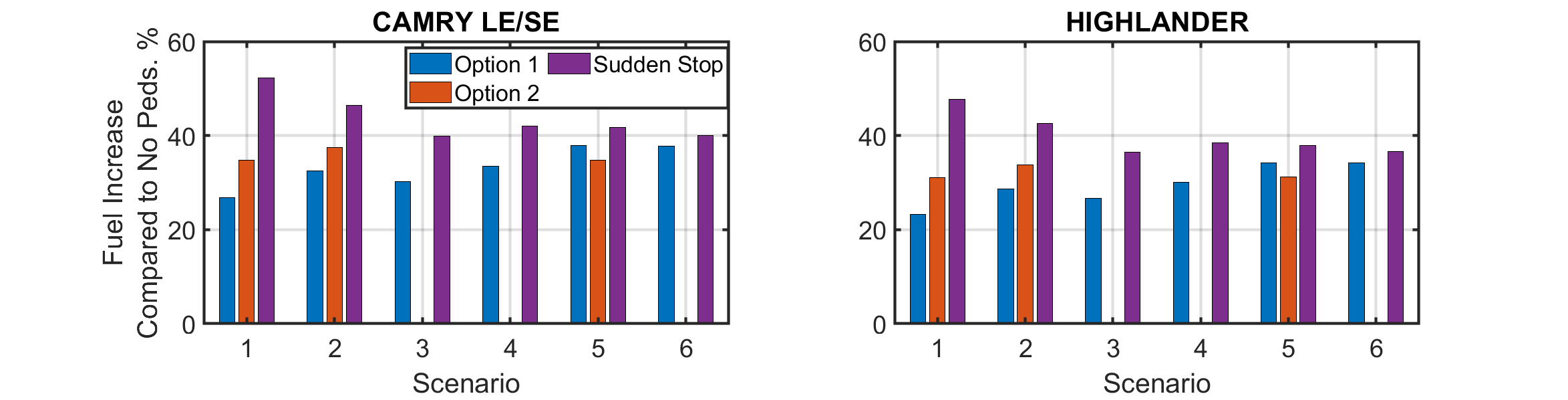}
         \caption{}
         \label{fig:total-fuel}
     \end{subfigure}
     \begin{subfigure}[b]{1\textwidth}
        \centering
        \includegraphics[width=\textwidth]{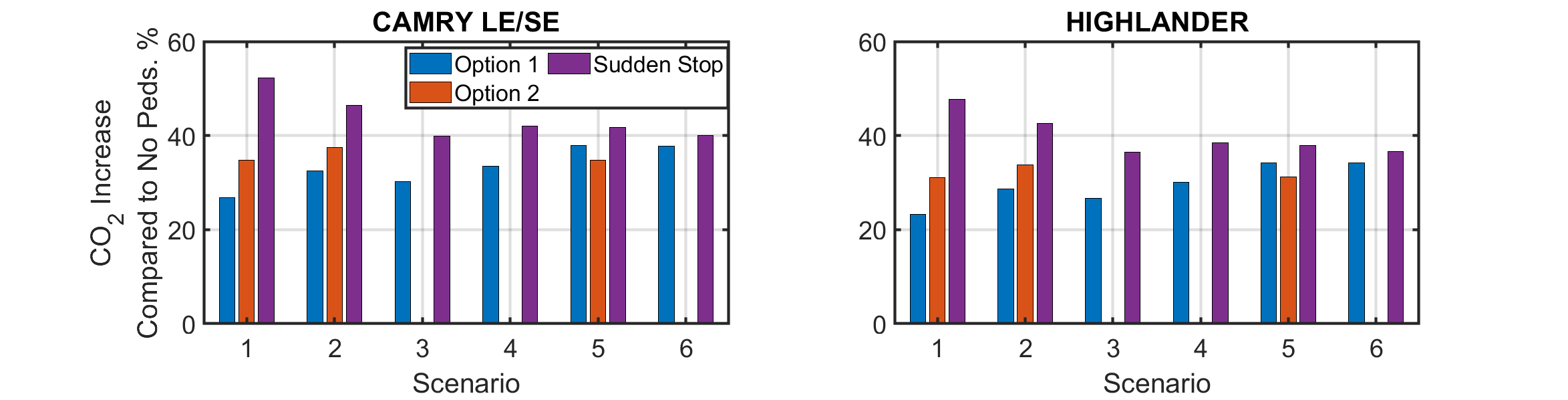}
        \caption{}
        \label{fig:total-co2}
     \end{subfigure}
    \caption{The increase of (a )fuel consumed and (b) emitted \ch{CO2} for all the trajectories compared to no pedestrian trajectory in all scenarios}
    \label{fig:increase}

\end{figure}

We measured the increase percentage of fuel as a result of midblock crossing compared to the case when there are no pedestrian (No Peds.). Fig.~\ref{fig:increase} shows that 
accommodating pedestrians in all scenarios increases the fuel consumption and emissions compared to the case where there are no pedestrians (No Peds.). However, receiving timely 
informative messages that allow the car to maintain a safe speed consumes less fuel and emits less \ch{CO2} than the Sudden Stop. This is because when the car maintains a lower speed 
in advance until the pedestrians finish crossing, it consumes less fuel and consequently emits less \ch{CO2}. Conversely, when the car does not have timely information and 
suddenly stops at the crossing location, and later resume its normal speed after passing it, it consumes more fuel because of the higher speed and acceleration phases.

\begin{figure}
     \centering
     \begin{subfigure}[b]{1\textwidth}
         \centering
         \includegraphics[width=\textwidth]{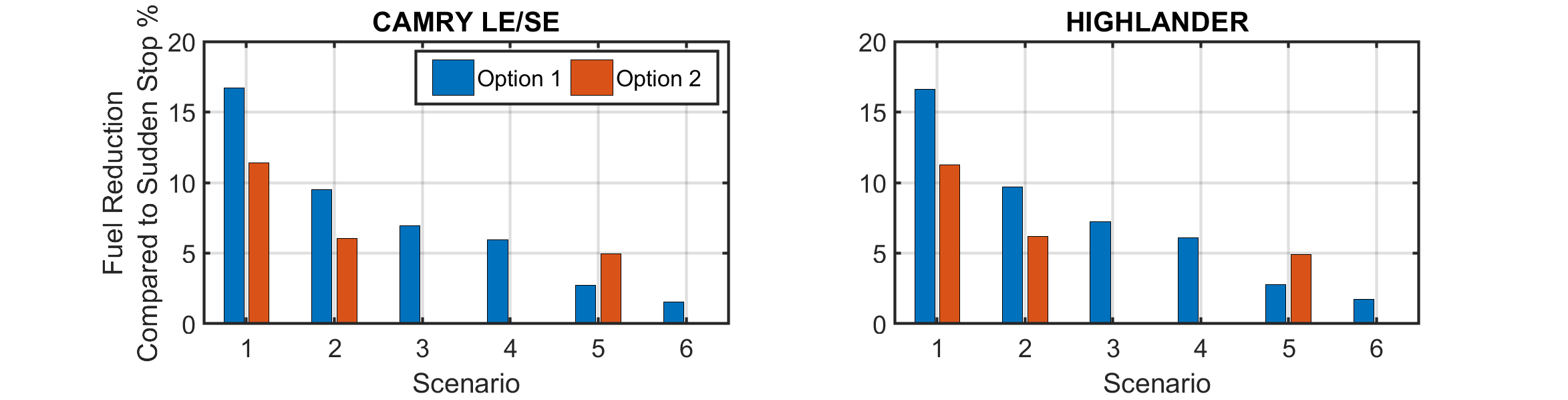}
         \caption{}
         \label{fig:fuelReduction}
     \end{subfigure}
     \begin{subfigure}[b]{1\textwidth}
        \centering
        \includegraphics[width=\textwidth]{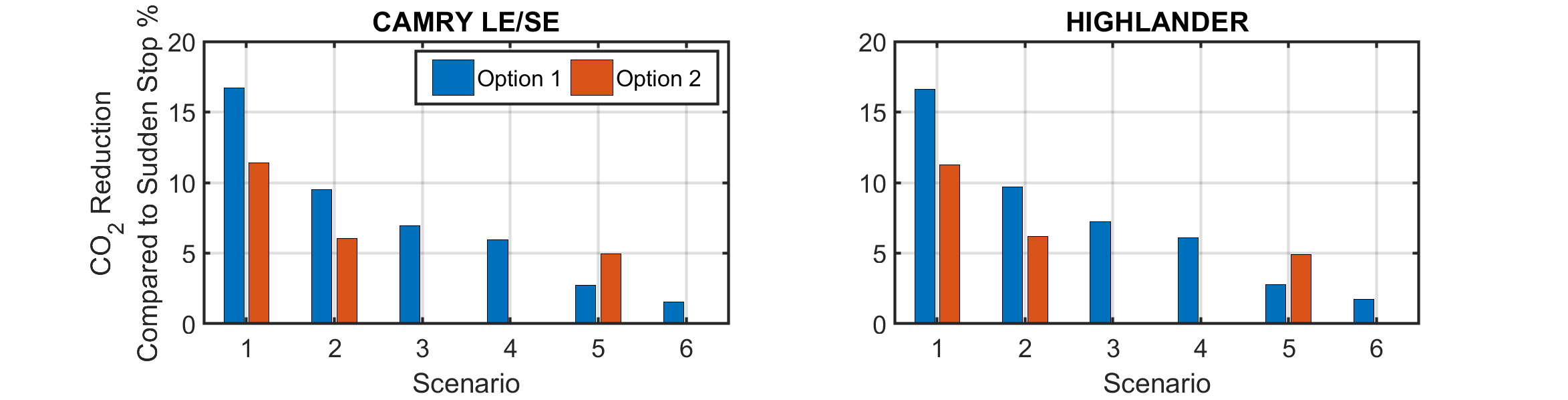}
        \caption{}
        \label{fig:CO2Reduction}
     \end{subfigure}
    \caption{The reduction of (a) fuel consumed and (b) emitted \ch{CO2} for all the trajectories compared to sudden stop trajectory in all scenarios}
    \label{fig:reduction}

\end{figure}

The reduction of fuel consumption and \ch{CO2} emissions in the proposed schemes is shown in Fig.~\ref{fig:reduction}. The reduction of Option 1 and Option 2 were compared 
with the (Sudden Stop) trajectory. As can be seen, the reduction in fuel consumption is higher in the two options. 
We can also see that Scenario 1 has the highest fuel and emission reduction.
This shows the effectiveness of the proposed schemes in this case. While in the other scenarios, as new information about pedestrian crossing at a closer location may require 
consuming more fuel as the time allowed to reduce the speed is less than in Scenario 1. 

To discuss the results in detail, we show duration of driving modes (acceleration, deceleration, cruising and idling) for all the trajectories in all the scenarios 
in Table.~\ref{tab:duration}. This data is applicable to the two cars since the speed profile is the same for both. From Table.~\ref{tab:duration}, we see that in Scenario 1 the 
two options consumed less fuel than sudden stop because in both trajectories the car maintained a lower speed than sudden stop. Another reason is that in sudden stop, and because 
the car does not have adequate information in advance, it tries to resume its normal speed after it passes the crossing location. This causes an increase in the acceleration 
duration, which consumes more fuel. This was also applied to the other scenarios.

\begin{table}[]
\centering
\resizebox{\textwidth}{!}{%
\begin{tabular}{clcccccc}
\multicolumn{1}{c}{{\color[HTML]{333333} Scenario}} &
  \multicolumn{1}{c}{Trajectory} &
  speed. $\mu$ [mph] &
  speed. $\sigma$ &
  Acceleration Duration [seconds] &
  Deceleration Duration [seconds] &
  Cruising   Duration [seconds] &
  Idling Duration [seconds] \\ \hline
                    & Option 1    & 23.934  & 7.3384   & 7    & 1.5 & 85   & 0    \\
                    & Option 2    & 23.6805 & 8.9587   & 7    & 1.5 & 86   & 0    \\
                    & No Peds     & 29.9982 & 1.07E-14 & 0    & 0   & 74.5 & 0    \\
\multirow{-4}{*}{1} & Sudden Stop & 23.6799 & 10.386   & 20.5 & 2.5 & 71.5 & 4.5  \\ \hline
                    & Option 1    & 23.2338 & 8.824    & 7    & 1.5 & 87.5 & 0    \\
                    & Option 2    & 23.1714 & 10.0558  & 7    & 1.5 & 88   & 0    \\
                    & No Peds     & 29.9982 & 1.07E-14 & 0    & 0   & 74.5 & 0    \\
\multirow{-4}{*}{2} & Sudden Stop & 23.0918 & 11.3951  & 16   & 2.5 & 78.5 & 13.5 \\ \hline
                    & Option 1    & 23.4776 & 8.6753   & 7    & 1.5 & 86.5 & 0    \\
                    & Option 2    & 23.4776 & 8.6753   & 7    & 1.5 & 86.5 & 0    \\
                    & No Peds     & 29.9982 & 1.07E-14 & 0    & 0   & 74.5 & 0    \\
\multirow{-4}{*}{3} & Sudden Stop & 23.4451 & 11.3236  & 15.5 & 2   & 77.5 & 14   \\ \hline
                    & Option 1    & 23.2619 & 9.4685   & 7    & 1.5 & 87.5 & 0    \\
                    & Option 2    & 23.2619 & 9.4685   & 7    & 1.5 & 87.5 & 0    \\
                    & No Peds     & 29.9982 & 1.07E-14 & 0    & 0   & 74.5 & 0    \\
\multirow{-4}{*}{4} & Sudden Stop & 23.3836 & 11.337   & 15   & 2.5 & 78   & 14   \\ \hline
                    & Option 1    & 22.6306 & 10.6188  & 7    & 2   & 90   & 0    \\
                    & Option 2    & 25.1097 & 8.0688   & 15   & 1.5 & 72.5 & 0    \\
                    & No Peds     & 29.9982 & 1.07E-14 & 0    & 0   & 74.5 & 0    \\
\multirow{-4}{*}{5} & Sudden Stop & 24.8747 & 8.9741   & 19.5 & 1.5 & 69   & 0    \\ \hline
                    & Option 1    & 22.5959 & 10.7654  & 7    & 1.5 & 90   & 0    \\
                    & Option 2    & 22.5959 & 10.7654  & 7    & 1.5 & 90   & 0    \\
                    & No Peds     & 29.9982 & 1.07E-14 & 0    & 0   & 74.5 & 0    \\
\multirow{-4}{*}{6} & Sudden Stop & 24.0758 & 10.6726  & 15.5 & 2   & 75.5 & 11.5 \\ \hline
\end{tabular}%
}
\caption{\em Speed statistics and duration of driving modes in the evaluated scenarios}
\label{tab:duration}

\end{table}

We conclude that receiving advance information about midblock crossing allows the car to reduce the fuel consumption by up to 16.7\% over the sudden reaction and on average 7.4\% for 
both options. Specifically, immediate deceleration reduced about  7.3507\% from the sudden reaction and deferred declaration reduced  7.4511\% on average. 
This conclusion applies to CAMRY data, and we achieved approximately similar reduction for the  HIGHLANDER model.  As the \ch{CO2} emission are highly correlated with fuel 
consumption, emissions were reduced correspondingly. It is worth mentioning that neither of the schemes affected the average speed for the whole trip which means that our scheme 
did not increase trip time.

\section{Concluding remarks}\label{sec:conc}
Our main contribution was to propose schemes that mitigate the environmental impacts (increased fuel consumption and \ch{CO2} emissions) of pedestrian midblock crossing by leveraging 
information about the location and expected duration of the crossing. 
Our extensive simulations showed that timely dissemination of pedestrian crossing information to approaching vehicles can reduce fuel
consumption and emissions by up to 16.7\% 


\bibliographystyle{unsrtnat}






\end{document}